\documentclass{mem}
\usepackage{natbib}\usepackage{txfonts}\usepackage{balance}
\usepackage{graphicx}
\usepackage[a4paper,breaklinks,dvipdfm]{hyperref}
\idline{75}{282}
\begin{document}
\def\teff{$T\rm_{eff }$}
\def\kms{$\mathrm {km s}^{-1}$}

\title{ 
TERZAN 5: the remnant of a pristine fragment of the Galactic Bulge?  
}

   \subtitle{}

\author{
B. \,Lanzoni\inst{1} 
\and the Cosmic-Lab Team\inst{1,2}
          }

  \offprints{B. Lanzoni; \email{barbara.lanzoni3@unibo.it}}

\institute{
Dipartimento di Fisica e Astronomia -- Universit\`a di Bologna, Viale Berti Pichat 6/2,
I-4027 Bologna, Italy
\and
http://www.cosmic-lab.eu
}

\authorrunning{Lanzoni}

\titlerunning{Terzan 5}

\abstract{Terzan\,5 is a stellar system in the Galactic bulge commonly
  catalogued as a globular cluster.  Through dedicated NIR photometry
  and spectroscopy we have discovered that it harbors two main stellar
  populations defining two distinct red clumps (RCs) in the
  colour-magnitude diagram, and displaying different iron content:
  [Fe/H]$\simeq -0.2$ and [Fe/H]$=+0.3$ for the faint and the bright
  red clumps, respectively.  In addition, a third minor population
  with significantly lower metallicity ([Fe/H]$=-0.79$) has been
  recently detected, thus enlarging the metallicity range covered by
  Terzan 5 to $\Delta$[Fe/H]$\sim 1$\,dex.  This evidence demonstrates
  that, similarly to $\omega$\,Centauri in the Galactic halo,
  Terzan\,5 is not a genuine globular cluster, but a stellar system
  that experienced a much more complex star formation and chemical
  enrichment history. Moreover the striking chemical similarity with
  the bulge stars suggests that Terzan\,5 could be the relic of one of
  the massive clumps that contributed (through strong dynamical
  interactions with other similar sub-structures) to the formation of
  the Galactic bulge.

 }
\maketitle{}

\section{Introduction}

\begin{figure}[t!]
\resizebox{\hsize}{!}{\includegraphics{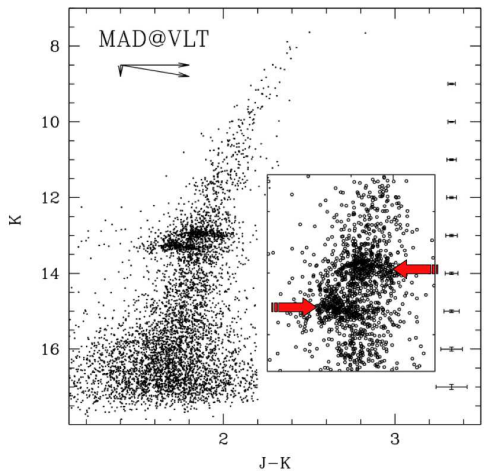}}
\caption{\footnotesize ESO-MAD ($K,\,J-K$) CMD of the central
  ($1'\times 1'$) region of Ter\,5.  The inset show a zoom in the RC
  region with the two distinct RCs clearly visible. The reddening
  vector is also plotted (from F09).}
\label{mad}
\end{figure}

Galactic Globular clusters (GCs) are dynamically active systems
\citep[e.g.,][]{fe09_bss,fe12}, old enough
\citep[e.g.][]{marinfranch09} to have witnessed the entire
evolutionary history of the Milky Way.  Increasing evidence of
multiple evolutionary sequences and significant spreads in the
abundance of a few light-elements is accumulating
\citep[e.g.,][]{piotto09,carretta10} and it suggests that GC formation
may have been more complex than previously thought. However the
striking homogeneity in the GC iron content indicates that these
systems still are the best approximations in nature of simple stellar
populations.  

The only clear exception known until recently was $\omega$\,Centauri
in the Galactic halo.  A variety of stellar populations \citep{lee99,
  pancino00, fe04, sollima07} showing a large ($>1$\,dex) spread in
the iron abundance \citep{nor95,sol05,johnson10,pancino11} have been
observed in this object, which is now considered to be the remnant of
a disrupted dwarf galaxy accreted by the Milky Way. The detailed
investigation of $\omega$\,Centauri can therefore provide us with
crucial information about the formation and evolutionary history of
our galaxy.  

Analogous findings in the bulge have been hampered by the severe
reddening conditions of this Galactic region.  Recently, however, we
have discovered that an object harbouring multi-iron populations
orbits the bulge \citep[][hereafter F09]{fe09}: it is named Terzan\,5
and it was commonly catalogued as a GC.  The results obtained to date
suggest, instead, that Terzan\,5 is the remnant of a much more massive
structure that contributed to form the Galactic bulge.  Here we
describe the main findings obtained to date about Terzan\,5, and the
overall scenario that is emerging \citep[all results are from
  F09,][]{lanz10,origlia11,origlia13,massari13}.

\section{The discovery}
Terzan\,5 is located in the inner Bulge of our Galaxy, in a region
affected by large and differential reddening: the average color excess
is $E(B-V)=2.38$ \citep{ortolani96,barbuy98,valenti07} and its
variation within the field of view covered by Terzan\,5 can be as
large as $\Delta E(B-V)\simeq 0.7$ \citep{massari13}. In the recent
past it received special attention because of its exceptionally large
population of millisecond pulsars, which amounts to $\sim 25\%$ of the
entire sample of these objects known to date in Galactic GCs
\citep{ransom05}.

By using near-IR high-resolution images obtained with the ESO-MAD, we
discovered (F09) that Terzan\,5 harbors two stellar populations
defining two distinct red clumps (RCs; see Fig. \ref{mad}): a
bright-RC at $K = 12.85$, and a faint-RC at $K = 13.15$, the latter
having a bluer ($J-K$) color.  Prompt spectroscopic observations (with
NIRSPEC@Keck) of a few stars in the two clumps have demonstrated that
the two populations have the same radial velocity (hence they both
belong to Terzan\,5), but a significantly different iron abundance:
[Fe/H]$\simeq -0.2$ for the faint-RC, and $+0.3$ for the bright-RC.
This makes Terzan \,5 the first GC-like system with multi-iron
sub-populations ever discovered in the Galactic Bulge (F09).
 
\section{Spectroscopic follow-up}
A NIRSPEC spectroscopic follow-up of 33 giants confirmed that the two
populations have different iron content and revealed intriguing
abundance patterns \citep{origlia11}.

\begin{figure*}[t!]
\resizebox{\hsize}{!}{\includegraphics{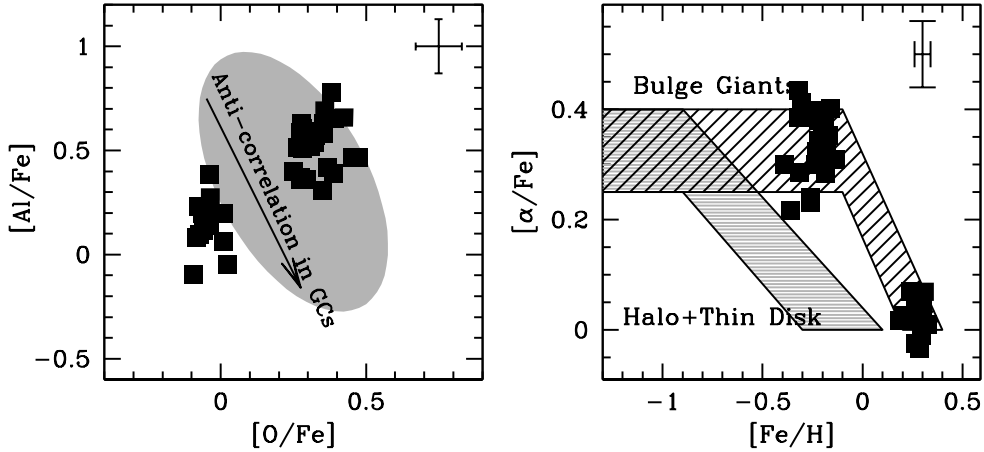}}
\caption{\footnotesize {\it Left panel:} Trend of [Al/Fe] {\it versus}
  [O/Fe] for 33 giants observed in Terzan\,5 \citep[black squares, from][]{origlia11}. It
  appears to be {\it orthogonal} to the anti-correlation (grey region)
  observed in genuine GCs (Carretta et al. 2009). {\it Right panel:}
  [$\alpha$/Fe] {\it versus} [Fe/H] for a sample of 33 giants in
  Terzan\,5 \citep[black squares, from][]{origlia11}, compared to
  those of the Galactic bulge and halo+thin disk giants (shaded and
  grey regions, respectively; e.g., \citealp{alves10}). The chemistry
  of the Terzan\,5 populations is consistent with that of the bulge
  field giants.}
\label{fig_origlia11}
\end{figure*}

First, the population as a whole, and also the two sub-components
separately, display a trend between [O/Fe] and [Al/Fe] which seems to
be {\it orthogonal} to the anti-correlation commonly found in genuine
GCs (left panel in Fig. \ref{fig_origlia11}).  Moreover, the two
populations show a very small spread ($\sim 0.1$ dex) in both [O/Fe]
and [Al/Fe], never exceeding the $1\sigma$ measurement errors, again
at odds with what found in GCs of any metallicity \citep{gratton04}.

Second, the overall iron abundance and the amount of
$\alpha-$enhancement of the two Terzan\,5 components (right panel
Fig. \ref{fig_origlia11}) suggest that the faint-RC component likely
formed from a gas mainly enriched by Type II supernovae (SNeII) on a
short timescale.  The larger [$\alpha$/H] ratio of the metal-rich
population indicates that it was additionally enriched from SNII
ejecta. Moreover, its about solar [$\alpha$/Fe]$=+0.03 \pm 0.04$
indicates that the progenitor gas was also polluted by SNeIa on longer
timescales.

\section{A third population discovered}
In the context of an ongoing radial velocity survey of Terzan\,5, we
found preliminary indications of the presence of a minor ($\sim 3\%$)
stellar population significantly more metal-poor than the faint-RC
component of Terzan\,5.  We therefore acquired high resolution spectra
of 3 such candidate metal poor giants on June 17, 2013 by using
NIRSPEC at Keck II, and we measured the chemical abundances of iron,
$\alpha$-elements, carbon and aluminum \citep{origlia13}. Based on the
measured radial velocity, the three stars turned out to be cluster
members. Their average iron abundance is [Fe/H]$=-0.79\pm 0.04$
r.m.s., significantly lower (by a factor of $\sim$ 3) than the value
of the faint-RC component ([Fe/H]$=-0.25$). This clearly points
towards the presence of a third, distinct population in Terzan\,5, and
it significantly enlarges the metallicity range covered by this
stellar system, which now amounts to $\Delta$[Fe/H]$\sim 1$\,dex.

\begin{figure}[t!]
\resizebox{\hsize}{!}{\includegraphics{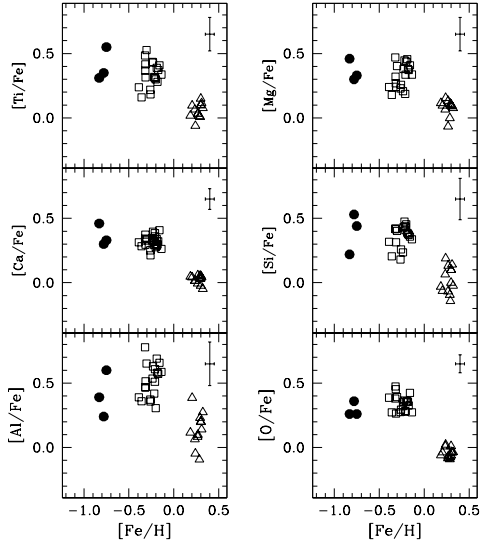}}
\caption{\footnotesize [$\alpha$/Fe] and [Al/Fe] abundance ratios as a
  function of [Fe/H] for the 3 newly discovered metal-poor giants
  \citep[solid dots;][]{origlia13}. The abundances measured for 20
  faint-RC (open squares) and 13 bright-RC (open triangles) giants
  \citep[from][]{origlia11} are also shown for comparison. Typical
  errorbars are plotted in the top-right corner of each panel.}
\label{fig_origlia13}
\end{figure}

As shown in Figure \ref{fig_origlia13}, the newly discovered
metal-poor population has an average $\alpha$-enhancement
[$\alpha$/Fe]$= +0.36\pm 0.04$, which is similar to that of the
faint-RC one. This indicates that both populations likely formed early
and on short timescales from a gas polluted by type II SNe.  As the
stars belonging to the faint-RC component, also these other giants
with low iron content show an enhanced [Al/Fe] abundance ratio
(average [Al/Fe]$= +0.41\pm 0.18$ r.m.s.) and no evidence of Al-Mg and
Al-O anti-correlations, and/or large [O/Fe] and [Al/Fe] scatters,
although no firm conclusion can be drawn with 3 stars only.

\section{The emerging scenario}
An intriguing scenario is emerging from these observational facts.

First, Terzan\,5 is not a genuine GC (as commonly thought), nor it can
derive from the merging of two globulars. Instead, it must have
experienced a complex star formation and chemical enrichment history.

\begin{figure}[t!]
\resizebox{\hsize}{!}{\includegraphics{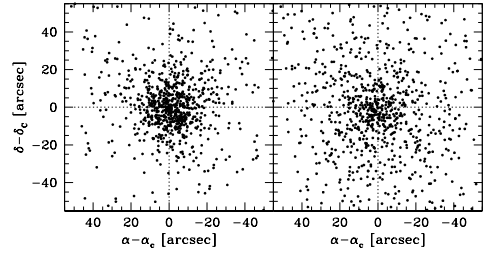}}
\caption{\footnotesize Map of the bright- and faint-RC populations
  (left and right panel, respectively), with the star coordinates
  referred to the gravity centre of Terzan\,5 \citep[from][]{lanz10}.}
\label{radist}
\end{figure}

Moreover, the evidence that the metal-richer component is
significantly (at $> 3.5\sigma$ level) more centrally concentrated
than the faint-RC population \citep[Fig. \ref{radist}; F09,][]{lanz10}
is a strong hint of self-enrichment.  In turn, this implies that
Terzan\,5 must have been much more massive in the past than today
\citep[its current mass being $\sim 10^6 M_\odot$;][]{lanz10}, thus to
be able to retain the high-velocity gas ejected by violent SN
explosions, from which the iron-rich stars populating the bright-RC
could form.

The exceptionally high metallicity regime of the two main stellar
populations found in Terzan\,5 also suggests a quite efficient
enrichment process through a large number of SNeII. These should have
also produced a large population of neutron stars, mostly retained
into the deep potential well of the massive {\it proto}-Terzan\,5.
The structural parameters recently re-determined for this system
\citep{lanz10} confirm that it has the largest collision rate among
all Galactic GCs. Hence, the formation of binary systems containing
neutron stars could have been largely favored, and it could have
promoted the re-cycling process responsible for the production of the
large population of millisecond pulsars now observed in Terzan\,5.

Finally, the chemical abundance patterns observed in Terzan\,5 very
closely resemble those of the bulge stellar population, which shows a
metallicity distribution with two main peaks at sub-solar and
super-solar [Fe/H] \citep[e.g.,][and references
  therein]{zoccali08,hill11,johnson11,rich12,utt12}.  These bulge
stellar populations show [$\alpha$/Fe] enhancement up to about solar
[Fe/H], and then a progressive decline towards solar values at
super-solar [Fe/H]. Such a trend is at variance either with the one
observed in the thick disk, where the knee occurs at significantly
lower values of [Fe/H], and with the rather flat and about solar
[$\alpha$/Fe] distribution of the thin disk (see right panel of
Fig. \ref{fig_origlia11}). Chemical abundances of bulge dwarf stars
from microlensing experiments \citep[e.g.][and references
  therein]{cohen10,bensby13} also suggest the presence of two
populations, a sub-solar and old one with [$\alpha$/Fe] enhancement,
and a possibly younger, more metal-rich one with decreasing
[$\alpha$/Fe] enhancement with increasing [Fe/H].  Moreover, a small
fraction ($\sim 5\%$, similar to that discovered in Terzan\,5) of
metal-poor stars ([Fe/H]$\sim -1$) has been detected also in the bulge
\citep[e.g.,][and references therein]{ness13a,ness13b}.

Indeed, both the iron and the [$\alpha/$Fe] abundance ratios measured
in Terzan\,5 \citep[see Figs. \ref{fig_origlia11} and
  \ref{fig_origlia13};][]{origlia11, origlia13} show a remarkable
similarity with those of the Galactic bulge stars.  {\it This strongly
  suggests that Terzan\,5 formed and evolved in deep connection with
  its present-day environment (the bulge) and it was not accreted from
  outside the Milky Way (as it seems to be the case for
  $\omega$\,Centauri).}  Within a self-enrichment scenario, the narrow
peaks in the metallicity distribution of the three Terzan\,5
populations can be the result of a quite bursty star formation
activity in the massive proto-Terzan\,5. However, Terzan\,5 might also
be the result of an early merging of fragments with sub-solar
metallicity at the epoch of the bulge/bar formation, and with younger
and more metal-rich sub-structures following subsequent interactions
with the central disk.  The current view
\citep[e.g.][]{korken04,immeli04,shen10} for the formation of a bulge
structure suggests a range of physical processes that can be grouped
in two main scenarios: (1) rapid formation occurring at early epochs
(as a fast dissipative collapse, mergers of
proto-clouds/sub-structures, evaporation of a proto-disk, etc.),
generating a spheroidal bulge populated by old stars, and (2)
evolution of a central disk/bar and its possible interaction with
other sub-structures on a longer timescale. 

Within this framework, Terzan\,5 might well be the relic of a larger
sub-structure that lost most of its stars, probably because of strong
dynamical interactions with other similar systems at the early epoch
of the Galaxy formation, and/or later on with the central
disk/bar. While most of the early fragments dissolved/merged together
to form the bulge, for some (still unclear) reasons Terzan\,5 survived
the total disruption. Hence, precisely deciphering the history of
Terzan\,5 would open new perspectives on our comprehension of the
formation and evolution of the Galactic bulge, and of galactic
spheroids in general.

\begin{acknowledgements}
The research is also part of the project {\it COSMIC-LAB}
(http://www.cosmic-lab.eu) funded by the {\it European Research
  Council} (under contract ERC-2010-AdG-267675).
\end{acknowledgements}

\bibliographystyle{aa}

\end{document}